# Studies of resistance switching effects in metal/YBa$_2$Cu$_3$O$_{7-x}$ interface junctions

A.Plecenik[1], M.Tomasek[1], T.Plecenik[1], M.Truchly[1], J.Noskovic[1], M.Zahoran[1],

T.Roch[1], M.Belogolovskii[2], M.Spankova[3], S.Chromik[3], and P.Kus[1]

[1] *Department of Experimental Physics, FMPI, Comenius University, 84248 Bratislava, Slovak Republic*

[2] *Donetsk Institute for Physics and Engineering, NASU, 83114 Donetsk, Ukraine*

[3] *Institute of Electrical Engineering, SAS, 84104 Bratislava, Slovak Republic*

**Abstract**

Current-voltage characteristics of planar junctions formed by an epitaxial c-axis oriented YBa$_2$Cu$_3$O$_{7-x}$ thin film micro-bridge and Ag counter-electrode were measured in the temperature range from 4.2 K to 300 K. A hysteretic behavior related to switching of the junction resistance from a high-resistive to a low-resistive state and vice-versa was observed and analyzed in terms of the maximal current bias and temperature dependence. The same effects were observed on a sub-micrometer scale YBa$_2$Cu$_3$O$_{7-x}$ thin film – PtIr point contact junctions using Scanning Tunneling Microscope. These phenomena are discussed within a diffusion model, describing an oxygen vacancy drift in YBa$_2$Cu$_3$O$_{7-x}$ films in the nano-scale vicinity of the junction interface under applied electrical fields.

## 1. Introduction

It is expected that the miniaturization and acceleration of silicon-based electronic memories reaches its limit within a decade. Typical dimensions of the active zone in a memory cell are tens of nanometers nowadays. After extrapolating it may shrink to a few nanometers in the next decade. In such a scale, regularly used voltages will generate strong electric fields on contacts. This can lead to tunneling of electrons or even migration of charged sub-nanometer particles. The process will cause dramatic changes in the active zone composition and may even lead to a non-functionality. This apprehension motivated a great deal of activities in the search for new technologies based on effects emerging in the nanometer scale. Their key feature should be the reversibility of the induced change.

The resistive switching (RS) effect observed in capacitor-like metal/insulator/metal junctions belongs to the most promising candidates for such a technology [1-4]. It is based upon a sudden change of a junction resistance caused by the electric field applied to the metal electrodes. This effect was observed in large variety of materials such as cuprates [5-6], titanates [7-9], some organic compounds [10-12], but mainly on transition-metal oxides [3-4,13-15]. It is expected that some of these materials can form a base for a resistive random access memory (ReRAM). Comparing with the magnetic RAMs, the ReRAMs can avoid problems with magnetic disturbs and allow fully electronic programming. Also, due to the nature of the RS effect, it is believed that the miniaturization of such ReRAM devices will be possible down to the nanometer scale. Although the mechanism of the RS phenomenon may vary for different oxide materials, in most cases it is believed to be related to a redistribution of oxygen vacancies. In the case of devices based on simple two-component oxides, the RS is usually unipolar (it can be switched to both "low-resistive state (LRS)" and "high resistive state (HRS)" within the same electric field polarity) and it is assumed to be caused by a creation and annihilation of conductive channels with high concentration of oxygen vacancies [13]. On the other hand, more complex transition-metal oxides usually exhibit a bipolar RS (switching to LRS and HRS states occurs at the opposite electric field polarity). It is most likely caused by the local change of the resistance within a nanometer-scale region in the oxide/metal electrode interface due to the modification of oxygen vacancies concentration [16]. A lot of effort has been done for the confirmation of the driving mechanism in the perovskite-type manganites and superconducting cuprates. The materials of the second type more convenient for explanation of the switching mechanism since their properties, including the influence of oxygen vacancies concentration, are well known thanks to extensive prior research of high temperature superconductivity [17].

In this work, we present our results of current-voltage (I-V) measurements for c-axis oriented Ag-YBa$_2$Cu$_3$O$_{7-x}$ (YBCO) planar junctions exhibiting a bipolar resistance switching from the HRS to the LRS and vice-versa. On this type of YBCO films, I-V characteristics exhibiting the RS phenomena were measured by cryogenic Scanning Tunneling Microscope (STM) on PtIr-YBCO point contact junctions as well. The STM measurements provide clear evidence that the effect discussed can also be observed in sub-micrometer scale junctions. Therefore it is suitable for possible applications in the next generation of non-volatile RAMs.

## 2. Experimental

High-quality c-axis oriented YBCO films with a thickness of about 200 nm were fabricated on LaAlO$_3$ (001) single-crystalline substrates by dc magnetron sputtering from a stoichiometric ceramic target. The deposition was performed in an oxygen atmosphere at a pressure of 240 Pa, a magnetron power of 4.5 W/cm$^2$ and a sample heater temperature of 815 °C. After that, the samples were annealed in oxygen (10$^4$ Pa) at 500 °C during 30 minutes with a subsequent cooling down to room temperature at a rate of 15 °C/min. Resistance vs. temperature dependences of such an YBCO thin film exhibited metallic dependences and the T$_{c0}$ at about 85 K (Fig. 1).

An optical lithography and Ar ion beam etching was then used to fabricate 100 μm wide YBCO strips. Subsequently the strips were planarized by SiO$_2$ (deposited using vacuum evaporation before a photoresist removal). After this step the photoresist was removed and 110 nm thick Ag counter-electrodes across the YBCO strip were fabricated with the lift-off optical lithography and vacuum evaporation. The width of the Ag counter-electrodes was 5, 10, and 15 μm. The detail of the final structure is shown in Fig. 2.

Transport properties of the prepared junctions were measured by a standard four-contact setup in a current-biased mode with Keithley 220 Current Source and Keithley 2000 Multimeter controlled by a computer via GPIB interface. The measurements were performed in the temperature range from 4.2 K to 300 K in a transport He Dewar container. In all measurements, the bottom YBCO electrode was biased and the Ag counter-electrode was grounded.

For STM measurements, a cryogenic STM from NT-MDT Company was used. On this device, the voltage bias between the STM tip and the sample was adjusted within +/- 10 V range and the corresponding current was measured up to +/- 50 nA. For all STM measurements, a PtIr wire was used as a STM tip. Similarly to the case of the planar junctions, the YBCO film was biased and the STM tip was grounded (Fig. 3). Based on several articles dealing with the STM measurements of YBCO [18-19] we assume that the STM tip was in direct contact with the YBCO surface due to the existence of a degraded insulating layer which served as a tunneling barrier instead of the sample-tip interspace. This causes the bad quality of the topographic image obtained with the STM (see the inset in Fig. 3) which is probably a combination of surface topography and inhomogenities of the insulating barrier formed on the YBCO surface. The measurements were done in the Oxford Instruments Optibath SXM cryostat down to liquid nitrogen temperature (77 K).

## 3. Results and Discussion

Typical I-V characteristics measured on the planar Ag-YBCO junctions and STM PtIr-YBCO point contact junctions exhibiting a hysteretic behavior related to the resistance switching from the HRS to the LRS and vice-versa are shown in Fig. 4a and 4b, respectively. The virgin state of the junctions was always a high-resistive one. Starting the measurements with a positive-polarity bias, no hysteresis was observed. Such a behavior is in agreement with the well known fact that a nanometer-scale layer of the YBCO in the junction interface vicinity is usually degraded (i.e., a high concentration of oxygen vacancies is already present and this part of the YBCO is in the insulating state [18-20] and hence is in the HRS). At the positive polarity the electric field forces more oxygen vacancies to enter the interface layer, but probability of their migration in this direction is very small due to already high concentration of vacancies. That is why the junction resistance is not changed considerably. Only for the negative biases and with a sufficient voltage stress the junction switches from the HRS to the LRS (point A in Fig. 4a) and remains in the LRS until a sufficient voltage stress of the positive polarity is applied (point B in Fig. 4a). We assume that at the switching voltage bias (points A and B) the electric field is strong enough so that the oxygen vacancies can overcome their activation energy and thus causing their migration from (switching to the LRS) or to (switching to the HRS) the interface layer. This model was strongly confirmed by measurements of the densities of states in the LRS and the HRS [16]. The switching points are not visible on the I-V curves measured by the STM due to the 50 nA current measurement limit.

Further, we have investigated the temperature dependence of the LRS and the HRS. The resistance was calculated from the voltage at the current of 10 µA in the case of the planar junctions and 50 nA in the case of the STM point contact junctions. For both types of the contacts, the HRS resistance is relatively stable at a fixed temperature, i.e. independent on the maximal value of the applied voltage or current, respectively. The LRS resistance depends on the maximal value of the applied current (for the planar junctions) or maximal value of the applied voltage (for the STM point contact junctions) at the negative polarity. In both cases, higher maximal negative bias voltage (in the case of the point contacts) or maximal negative bias current (in the case of the planar contacts) applied on the junctions results in lower resistance of the LRS (Fig. 5). In our opinion, such a behavior means that the oxygen vacancies are still present in the YBCO interface layer even in the LRS state, i.e. the lowest possible resistance of the LRS has not yet been reached for the maximal bias voltage.

The R-T characteristics of the Ag-YBCO planar contacts in both states, the HRS and the LRS, are shown in the inset of Fig. 5a. It clearly shows that the resistance of the HRS is considerably increasing with the decreasing temperature what is usual for an insulating barrier. If we have Metal-Insulator-Metal (or Metal-Insulator-Superconductor) junction, the R-T characteristics of such system depend on the quality of the insulator used as tunneling barrier. In real tunneling junctions the tunneling resistance usually exponentially increases with decreasing of the temperature due to excitation states in the barrier. The situation in YBCO is much more complicated. One of the reasons is the YBCO phase diagram. If we have depleted layer of YBCO in vicinity to YBCO/metal interface the YBCO can exhibit metallic behavior at 300 K. But at lower temperature its part can undergo to the antiferromagnetic insulator. This phenomenon can give additional reason for increasing of the tunnel junction resistance with decreasing of the temperature. In the LRS the YBCO interface layer is in all cases more doped than in HRS and all above mentioned phenomena are more or less absent. In this case we do not expect rapid changes of the tunnel junction resistance with decreasing of the temperature. It was not possible to make such STM measurements with a stable contact over all the temperatures. Therefore we measured the I-V characteristics (and thus switched the junction from the HRS to the LRS and back) within the fixed voltage range (-8 V to 8 V) for every temperature separately and extracted the resistance of the HRS and the LRS from the I-V curves at 50 nA. Similarly as for the planar junctions, the resistance of the HRS was increasing with the decreasing temperature, while the resistance of the LRS does not change considerably with temperature variation as one can see in Fig. 6b. From the I-V curves (in the same figure) it is clear that the strength of the barrier formed by the YBCO interface layer is increasing with the decreasing temperature. Similarly as in the case of planar junctions, this is most probably caused by two factors: a lowering of the barrier transparency due to less thermal excitations from embedded charge traps and a gradual transition of the interface layer into the antiferromagnetic and more insulating phase according to the YBCO phase diagram.

Further, we have observed that the nature of the switching in the planar contacts is continuous at room temperature (Fig. 4a), and it is changing to a discrete step-like switching at low temperatures (Fig. 6a). Surprisingly, also at room temperature this gradual change occurs only above a well defined threshold voltage value. Its temperature dependence is shown in the inset in Fig. 6a. The absolute value of the threshold voltages increases with the decreasing temperature, while the negative "set" threshold voltage (the HRS to the LRS switching) is always higher than the positive "reset" voltage (the LRS to the HRS switching). The temperature dependence of the threshold voltages is non-trivial with a change of slope at about 200 K. This is most probably caused by a change of oxygen migration mechanism due to oxygen ordering in CuO chains and consequent increase of the activation energy for oxygen motion [17]. It was not possible to perform RS below 100 K on the Ag-YBCO planar junctions as the current corresponding to the threshold voltage destroyed the micro-structure. However, it was still possible to switch the junction above 100 K and than cool it down in LRS or HRS to study properties of both states down to 4.2 K.

4. Conclusions

Current-voltage characteristics of planar Ag-YBCO junctions as well as STM PtIr-YBCO point-contact junctions have been measured in the temperature range from 4.2 to 300 K and from 77 to 300 K, respectively. In both cases, the hysteretic behavior related to the resistance switching from the HRS to the LRS and vice-versa was observed. We have shown that at the fixed temperature the resistance of the HRS is relatively stable while the resistance of the LRS was decreasing when the maximal current or voltage value was increased. The measured temperature dependences demonstrate that the switching threshold voltage, the strength of the tunneling barrier formed on the YBCO/metal interface in the HRS and the HRS resistance are increasing with the decreasing temperature. On the other hand, the properties of the LRS are less temperature-dependent.

We discussed the experimental data within a diffusion model, taking into account the oxygen vacancy drift in the YBCO film in the sub-micrometer scale vicinity of the junction interface under applied electrical fields. The assumed model is strongly supported by our previous results published in [16]. Moreover, we demonstrate with the STM measurement that the resistance switching can also be realized in the sub-micrometer scale junctions. This observation is important for the possible miniaturization of the ReRAM devices based on the effect.

## 5. Acknowledgements

The work was supported by the Slovak Research and Development Agency under the contracts VVCE-0058-07, APVV-0034-07, APVV-0432-07 and LPP-0176-09.

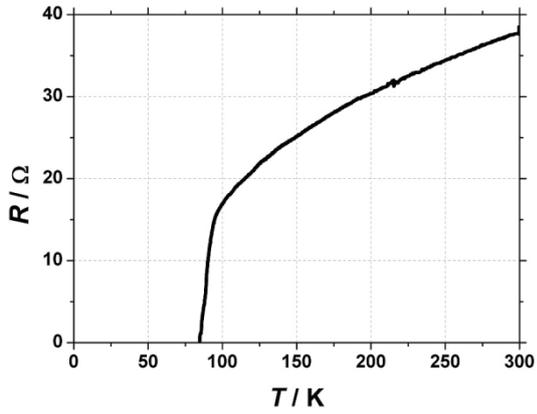

Fig. 1

R-T dependence of the YBCO thin film fabricated on LaAlO$_3$ single-crystalline substrates by dc magnetron sputtering.

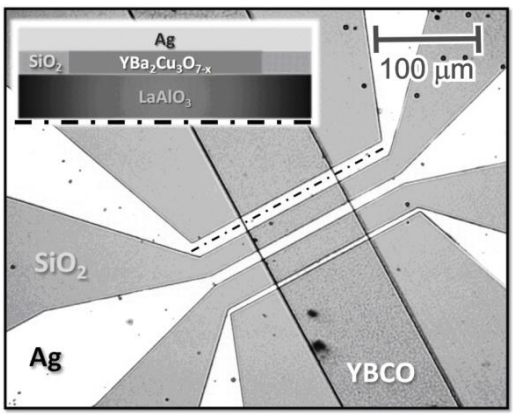

Fig. 2

Optical microscope photograph and cross-section schematic diagram (inset) of the Ag-YBCO planar junctions.

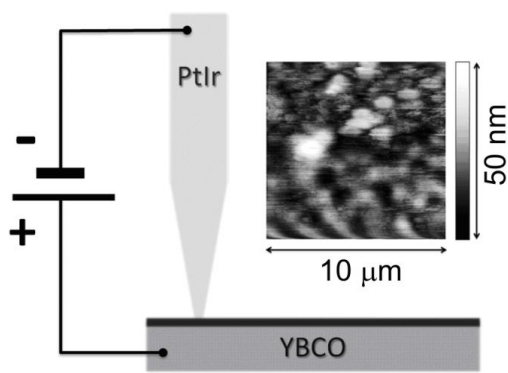

Fig. 3

Schematic diagram of the experimental setup for the STM measurements. Inset: Surface topography of the YBCO thin film measured by the STM in a constant current mode.

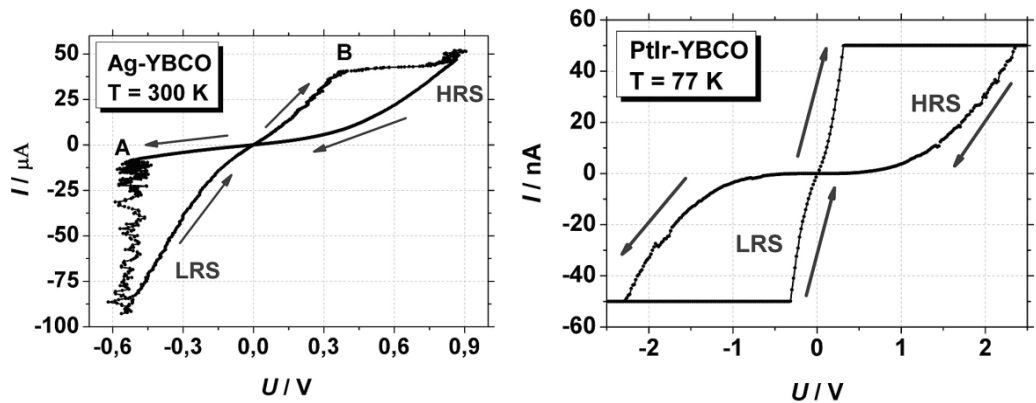

Fig. 4

Typical current-voltage characteristics of a) the Ag-YBCO planar junction at 300 K (point A shows where the junction starts to switch from the HRS to the LRS state and point B vice-versa) and b) the PtIr-YBCO STM tip junction at 77 K. The voltage was sweeped from -8 to +8 V and back.

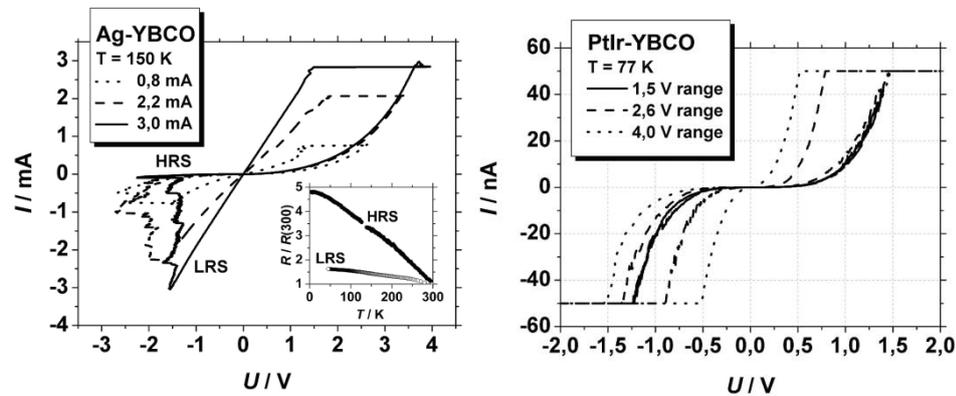

Fig. 5

a) Current-voltage characteristics of the Ag-YBCO contact for different current ranges with the temperature dependence of the LRS and the HRS shown in the inset.
b) Current-voltage characteristics of the STM PtIr-YBCO contact for different voltage ranges show the dependence of the LRS and the HRS on the maximal voltages applied.

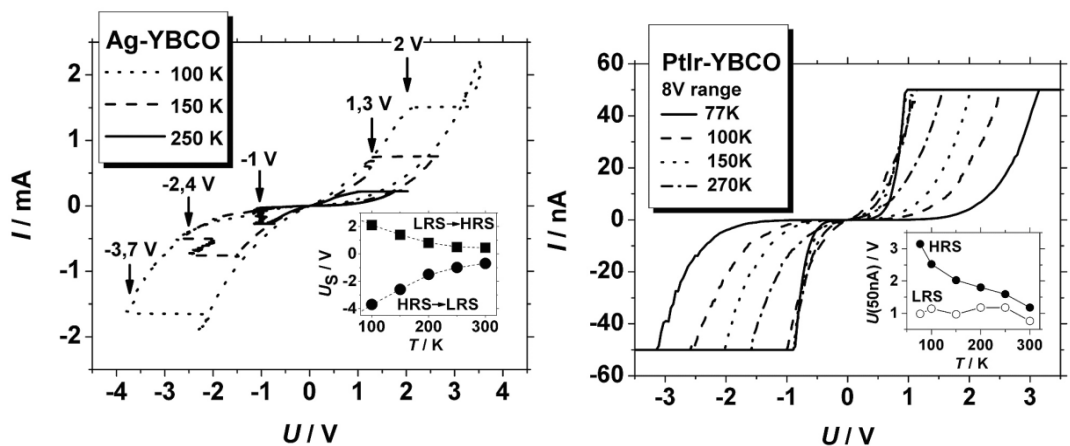

Fig. 6

a) Current-voltage characteristics of the planar Ag-YBCO contacts at 100, 150, and 250 K. The temperature dependences of threshold voltages are shown in the inset.

b) Current-voltage characteristics of the STM PtIr-YBCO contact at 77, 100, 150, and 270 K. The temperature dependences of the LRS and HRS are shown in the inset.